# Design of a multimedia processor based on metrics computation

Nader Ben Amor[1,2], Yannick Le Moullec[3], Jean Philippe Diguet*[1], Jean Luc Philippe[1] and Mohamed Abid[2]

[1]Lester Laboratory, Research Center, UBS University - BP 92116 - 56321 LORIENT Cedex Lorient, France.
[2]GMS Unit, Department of Electrical engineering, ENIS engineering school, B.W.P 3038 Sfax, Tunisia.
[3] Center for Embedded Software Systems (CISS) Fr. Bajers 7, B1-211 DK-9220 Aalborg, Denmark.

**\* corresponding author:** Jean-Philippe.Diguet@univ-ubs.fr  Fax number (+33) 2 97 87 45 27

## Abstract

Media-processing applications, such as signal processing, 2D and 3D graphics rendering, and image compression, are the dominant workloads in many embedded systems today. The real-time constraints of those media applications have taxing demands on today's processor performances with low cost, low power and reduced design delay.
To satisfy those challenges, a fast and efficient strategy consists in upgrading a low cost general purpose processor core. This approach is based on the personalization of a general RISC processor core according the target multimedia application requirements. Thus, if the extra cost is justified, the general purpose processor GPP core can be enforced with instruction level coprocessors, coarse grain dedicated hardware, ad hoc memories or new GPP cores. In this way the final design solution is tailored to the application requirements. The proposed approach is based on three main steps: the first one is the analysis of the targeted application using efficient metrics. The second step is the selection of the appropriate architecture template according to the first step results and recommendations. The third step is the architecture generation. This approach is experimented using various image and video algorithms showing its feasibility.

## 1- Introduction

We are currently experiencing an important increase in the use of embedded devices with powerful multimedia capabilities such as speech analysis and synthesis, character recognition, video compression, and graphics animation. Due to the various needs of such applications, embedded devices have to handle various data types and various complex tasks under hard real-time constraints. The need for real-time processing of complex algorithms is further accentuated by the increasing interest in other new domains like 3D image.
Especially in the domain of embedded systems, the main design constraint is the time to market priority since the availability of a new product at short time even not perfectly optimised is the key point for its commercial success. Another important point in this domain is the opportunity to take advantage of the application characteristics in order to optimize the energy / time / QoS tradeoff.
Thus, our strategy is to define a framework that provides a simple and fast analysis tool. The first point is to start with a typical software specification which is automatically transformed into a graph-based internal representation. The idea is to take as an input a standard code without performing any additional effort and to analyze it. Even if the false data dependencies have been eliminated, the resulting hierarchical graph still reflects the designer or standard authors point of view. If such a specification fits with a low cost, low power embedded processor, this is probably the more interesting solution. Secondly, our objective is to extract from the various granularity levels of this specification opportunities of parallelisms, which could be efficiently exploited on alternative architectures. The availability of parallelisms involves several tradeoffs factors.  The first one is the speed up of critical functions through resource allocation versus the area increase. This one can also means a significant static power growth [17]. Secondly time savings can be practically turned into power savings through the management of voltage / frequency couple and dynamic real-time scheduling, but it also implies a subsequent delay/power/area overhead.
Thus design methodologies are required to rapidly test and settle parameters such as the selection of instructions, the capabilities of local or I/O memories, the bandwidth of communication channels, the parallelism of processing units, the choice of dedicated hardware.
In our design space exploration strategy, a first step consisting in a metric-based analysis is performed rapidly without any architectural directive. In a second step the results are used to sketch the target architecture in order to perform a first set of estimations. The analysis of the metric results open different opportunities to compare alternative specifications or to propound the proper architecture style for a given (sub)function or task. These features include the wider/deeper trade-off, the ratio between explicit parallelism and the pipeline depth, the necessity of complex control instructions, the requirements in terms of local memories and specific bandwidth and the need of processing resources for address generation. This paper deals with these metrics, which are a key step to face the CAD challenge and rapidly converge towards the right design solution. The rest of the paper is organized as follows. Section 1.1 and 1.2 give an overview about respectively the use of metrics in the design space exploration and the different methods to design a multimedia processor. Section 2 details our approach. This paper focuses on the first step only. Section 3 details the internal graph-based representation used in our approach. Section 4 details the different metrics used in this paper for application specification. Section 5 shows experimental results for various multimedia applications. Finally, in section 6 we conclude about our work and present some perspectives.

## 1.1 State of the Art

This subsection present an overview of the different metrics used in various co-design approaches. It includes also an overview of various design methods of multimedia processors.

### 1.1.1 Metrics

Previous works dealing with metrics have been completed in the areas of high-level synthesis [9, 6] and hardware software codesign [9, 18, 19].

In [1] the metrics provide algorithm properties regarding a hardware implementation. The quantified metrics address the concurrency of arithmetic operations based on uniformed scheduling probabilities and the regularity that measures the repetition rate of a given pattern. In [6], some probability based metrics are proposed to quantify the communication link between arithmetic operators (through memory or registers). These metrics focus on a fine grain analysis and are mainly used to guide the design of datapaths, especially to optimize local connection and resource reuse. An interesting method for processor selection is presented in [5]. Three metrics representing the orientation of functions in terms of control, data transformation and data accesses orientation of functions are computed by counting specific instructions from a processor independent code. Then a distance is calculated, using specific characteristics of processors regarding their control, bandwidth and processing capabilities. Moreover the technique doesn't take into account instruction dependencies and there is no detail about the different kinds of access memory regarding the abstract processor model used. Finally, in [18] finer metrics are defined to characterize the *affinity* between functions and three kinds of targets: GPP, DSP and ASIC. The metrics result from the analysis and counting of C code instructions in order to highlight instruction sequences which can be DSP-oriented (buffer circularity, MAC operations inside loops, etc.), ASIC-oriented (bit level instructions) or GPP-oriented (conditional or I/O instructions ratio). Then a HW/SW partitioning tool is driven by the *affinity* metrics. Like [5] these metrics are dedicated to HW/SW partitioning, they don't exploit instruction dependencies and address a fixed (C procedures) granularity. Moreover, the locality of data bandwidth is not clearly taken into account.

### 1.1.2 Design of multimedia processor

To design a multimedia processor, many approaches can be considered. In [10] the authors propose a complete custom design of a processor (and its compiler), this method is very time consuming, since it doesn't use an IP core, and moreover doesn't achieve a low cost solution. Some other approaches based on IP cores are cost effective and require a reduced global design time. Some approaches add hardware modules (or coprocessors) to the original processor. Others add special multimedia instructions to the processor instruction set. In [11], a dedicated unit (called videocore) is added to the ARM processor to handle most of H.263 computationally intensive functions like motion estimation and DCT/IDCT transforms. In [12], a set of dedicated instructions is added to the ARM processor for multimedia operations. Those operations are typically special arithmetic operations, data manipulation like rearrangement and formatting. In [13], a processor dedicated to video compression is presented. Its control unit is divided on two hierarchical level: a high level unit (a RISC controller) which control execution of relatively simple operation like memory transfer, arithmetic and logic operation. Complex operation like those relative to motion estimation and DCT transform are controlled by a low level control unit. Data dependant operations (like VLC and VLD) are handled by a special hardware module. In [4] is presented the TANGRAM that is a coprocessor dedicated to scenes compositing at the display in the MPEG-4 decoding. Added coprocessors can be DSP. In [7] is presented the MVP (multimedia video processor) witch is based on a RISC processor coupled to four DSPs.

The approach defined in this paper differs from those previously described and try to offer a more general approach that includes an analysis step (using metrics) of the multimedia tasks used to efficiently select and specify software or hardware IP.

## 2 An approach for the design of a multimedia embedded processor

Figure 1 shows the proposed approach which is integrated in the Design Trotter environment [16]. This approach is based on two main steps which are described in what follows.

### 2.1 Metric computations of the target application

This step investigates the algorithmic complexity of the tasks to be implemented. It is used to analyze the algorithms without any consideration of the processor architecture.

For this purpose, various metrics have been defined. This step includes several substeps. The starting point for this first step is the application description written in the C language. This description may have different C functions. This description is automatically translated into a HCDFG graph.. The computation of the metric is based on this graph representation, which is detailed in section 3. We have implemented four of those metrics in Design Trotter: MOM, COM, HDRM and $\gamma$ metrics, these metric provide the memory/processing orientation, control orientation, the memory reuse and the average parallelism respectively.

For each function of the application and each level of graph hierarchy these metrics are computed. Then by analyzing those numeric results, we can classify the different functions that constitute the target application according to their behaviour. For instance, on the one hand control oriented functions with few parallelism opportunities would constitute promising candidate for a GPP software implementation. On the other hand, high parallelism functions (i.e with high $\gamma$ values) with few tests are candidate for hardware implementation.

An analysis of memory requirements of the target application is also performed. Indeed, memory modules dominate the cost, the performance and power consumption of embedded systems especially in image and video processing. Studying the impact of parallelism on memory size is important for trading off system performance against area cost. A memory bandwidth analysis is rapidly performed for each level of the graph hierarchy and the Balasa method [3] is currently adapted to our model in order to derive memory size optimization within loop nests.

### 2.2 Parameterization of the dedicated processor

This parameterization is performed according to the first step classification. It concerns the hardware IP (available or estimated) and the processor core dedicated specification.

Our approach consists in customizing existing processor architecture rather than creating an entirely new ASIP tuned to an application. For this, first generic dedicated models for multimedia application are defined. In this case study the generic architecture (see Figure 2) is based on a free IP SPARC LEON [8] that can be upgraded with co-processors: simple operators like multipliers, ALU, …, but also more complex functions such as DWT provided in the form of hardware IPs. The communication model is based on specific shared memories with generic hardware interface [14]. "Data in" and "data out" memories are defined for each hardware generic IP, if HW/HW communications are implemented a merging between "in" and "out" memories is performed [1].

Based on the characterisation data, three classes of functions are built. The first one includes candidates for hardware implementation, it can be for instance data-flow functions with sparse test operations (with low COM and high $\gamma$ values) . These will be added later as hardware IP to the LEON processor core [8] to obtain a low cost media processor. Those models include widely used modules that cover various multimedia applications like image transformation (DCT, wavelet transformer DWT, etc.) and classical image processing algorithm like filter operations, and also video processing functions like motion estimation and run-length coding. The second class contains typical software functions; it can be for instance control-dominated functions with few spatial parallelism (high COM and low $\gamma$ values). The third class of functions incorporates functions without a clear orientation, an advanced RTOS partitioning tool [1] will be used to perform a design space exploration.

Different implementation solutions feed the partitioning tools, the coarse grain options are based on HW accelerators estimations [21] or HW IP availability, the fine grain alternatives are based on the generic processor core with possibly instruction level coprocessors. Thus the metrics are used to guide the IP choice and specification but also to reduce the huge multi granularity design space.

Hardware IP are written using the VHDL language. Generic parameters of a video IP can be block size, macro-block size, motion threshold detection, filter window, and local memory size. The multimedia models are generic to allow maximum flexibility: their characteristic parameters can be tuned according to the application requirements.

Once this library is created, the next sub-step is to parameterise those dedicated modules according to the application specification (i.e. according to metrics results). Those parameters can be local memory size, image size, types of arithmetic operators, etc.

The parameterisation sub-step is followed by the automatic personalization of the multimedia modules VHDL specification. Software functions are scheduled to efficiently exploit the main processor resources.

# 3 Efficient graph-based specification

In this section, we detail the HCDGF graph-based representation of a C application. This representation is obtained automatically using a parser.

## 3-1 Definitions

Each C function of the specification is a node at the top level of the Hierarchical Control and Data Flow Graph (HCDFG). A function is a HCDFG. A HCDFG is a graph that contains only HCDFGs and CDFGs. A CDFG contains only elementary conditional nodes and DFGs. A DFG contains only elementary memory and processing nodes. Namely, it represents a sequence of non-conditional operations. There are three kinds of elementary (i.e., non-hierarchical) nodes of which the granularity depends on the architectural model: a processing node represents an arithmetic or logic operation. A memory node represents a data transfer (memory operation). Its parameters are the transfer mode (read/write), the data format and the hierarchy level that can be fixed by the designer. A conditional node represents a test operation (if, case, loops, etc.) There are also three types of dependencies represented by edges: a control dependency indicates an order dependency between operations without memory transfers (e.g., index computation before array access). Control dependency edges can also be used to impose an order between independent operations or graphs in order to favour resource usage optimization. A scalar data dependency between two nodes A and B indicates that node B uses a scalar issued from B vertex. A multi-dimensional data dependency is a data dependency where data produced is not a scalar but an array. Such an edge is created between a loop CDFG that reads an array produced by another loop CDFG.

## 3.2 Graph creation rules

The graph is travelled with a depth-first search algorithm. A HCDFG/CDFG is created when a conditional node is found at the next hierarchy level. When no more conditional nodes are found, a DFG is built. In order to facilitate the estimation process, CDFG patterns have been defined to rapidly identify *loop, if etc.* constructs. Another important point is that the model covers the complete application complexity. Thus, index computation (address computation), conditional tests and loop index evolution are represented with DFGs.

We distinguish several types of memory nodes:
1. input/ouput nodes (N1)
2. temporary data (produced by computations) (N2)
3. re-usable data (re-used input nodes) (N3)
4. accumulator data (N4)

N1 data are always global, N4 data are always local, N2 and N3 data can initially be local (stored in the register file) but they can be moved to the global memory if ever the local memory size becomes to small as compared to the application requirements. The first step of the metric calculation is located at the highest level of abstraction, without any architectural assumption. So, the data accesses considered are the global ones, corresponding to N1 data nodes.

A HCDFG example is depicted in Figure 3.

## 3.3 Hierarchical Characterization

The HCDFG representation enables multi-level granularity specification and characterization. Therefore the notion of function can correspond to several levels of granularity. At the lowest level, a function can represent for example a FIR filter. At an intermediate level, a function can represent a DWT. At the highest level, a function can represent a JPEG2K encoder. The scheme used for characterizing the application specification is based on a hierarchical bottom-up approach. The characterization results obtained for a certain level in the specification are combined together in order to characterize its upper level. The lower level characterization is performed with a fine grain granularity. At that level, the type of operations can be either processing (shifting, multiplications etc.) or data transfer. Once the lower levels have been estimated, the higher levels are estimated through combinations. This step can be performed rapidly as the information relevant to each low level function has been saved within its graph. Figure 3 shows a HCDFG specification example.

## 4- Metrics Computation

In this section, we define 4 metrics: $\gamma$ (Parallelism Upper Bound Metric), MOM (Memory Orientation Metric) and COM (Control Orientation Metric). We explain how they are computed for the leaf graphs and how they are combined to characterise CDFGs and HCDFGs.

## 4.1 γ metric

For a DFG graph γ is defined by formula (1) in Table 1. The critical path, noted CP, in a DFG graph, is the longest sequential chain of operations (processing, control, memory) expressed in terms of cycle number. CP is computed for each hierarchical level with a data and control dependency analysis. Our analysis method is not exclusively statistical contrary to [5] metrics. As defined, γ indicates the upper bound of spatial parallelism available at a given hierarchy level. For instance, if a HCDFG contains five parallel DFG where each DFG is fully sequential, then γ equals one for each DFG and five at the HCDFG level. The γ metric enables the classification of application functions according to their criticality, namely their capability to exploit the available parallelism. In the following design steps functions with highest γ can be first considered since they have the most important optimization potential regarding the acceleration and consequently energy savings. Note also, that it is also used to distribute cycle budgets to functions during the estimation and synthesis design steps.

Functions with high γ values can then be considered as appropriate to architectures with large explicit parallelisms. Functions that have a low γ value (circa 1) are rather sequential, so the acceleration can only be reached by exploiting temporal parallelism (i.e. deep pipeline).

## 4.2 combination rules

The metrics are computed in a bottom-up way, there are firstly calculated for leaf DFGs, then are computed for higher level CDFGs and HCDFGs with combination rules according to sequential, parallel, exclusive and loop structures. Hereafter are introduced combination rules for sequential, parallel, IF and FOR patterns for the computation of γ. The same approach is used for the other metrics.

A "IF" CDFG is composed of three subgraphs. The first one specifies the IF condition, the two others correspond to the true and false branches. For this graph, γ is calculated with the formula (2) in Table 1 where "Ptrue" and "Pfalse" are the probabilities to execute the true and false branches respectively. The branches are considered equiprobable by default but this can be modified after profiling the application. "Nopc" is the number of operations (global memory accesses and processing nodes) in the condition graph, "Noptrue/false" are the numbers of operations in conditional branches.

The computations of γ for "DO-WHILE" and "SWITCH" graphs are generalizations of "FOR" and "IF" formulas respectively. To determine the γ value of a HCDFG graph, we have to analyze its hierarchical structure. Figure 4 shows an example of a HCDFG composed of two nested "IF" CDFGs. The algorithm calculates Nop and CP of the following graphs: IF_DHeq0_CONDITION and IF_DHeq0_TRUE since they are simple DFGs and do not contain any subgraph. The IF_Dheq0_FALSE graph contains a subgraph (IF_TMPsup), therefore our algorithm goes down into the hierarchy, determines $Nop_{IF\_TMPsup}$ and $CP_{IF\_TMPsup}$ values. Then $Nop_{IF\_Dheq0\_FALSE}$ and $CP_{IF\_Dheq0\_FALSE}$ are computed. Finally, $Nop_{IF\_DHeq0}$ and $CP_{IF\_DHeq0}$ (and so $\gamma_{IF\_DHeq0}$) can be determined. This approach is recursively applied to the whole graph in order to compute its metric values. A HCDFG can be made of sequential and parallel graphs. For sequential graphs we use formula (3) in Table 1 to calculate γ. If there are parallel graphs (or combination of parallel and sequential graphs) we use formula (4) in Table 1.

## 4.3 MOM (Memory Orientation Metric)

MOM metric is defined by the general formula 5 in Table 1. MOM indicates the frequency of memory accesses in a graph. MOM values are normalized in the [0;1] interval. The closer to 1 MOM is, the more the function is considered as data-access dominated. Therefore in the case of hard time constraints, some high performance memories are required (large bandwidth, dual-port memory, etc.) as well as an efficient use of memory hierarchy and data locality [20]. To calculate MOM metrics, we follow the same approach as for γ computation. For a DFG graph, the global memory and treatment nodes are enumerated and saved as graph attributes. Then the MOM value is computed for the DFG. Those attributes are used to deduct MOM metrics for graphs located at higher hierarchical levels. More details about MOM computation is available in [15].

## 4.4 COM (Control Orientation Metric)

To calculate this metric, test operations, namely the following operators: <=, <, >, >=, !=, must be identified. COM is defined by the general formula (6) in Table 1. It indicates the appearance frequency of control operations (i.e, tests that can be eliminated at compilation time) in a graph. We follow the same approach as for γ calculations for different cases of CDFGs. Additional information about metrics calculation can be found in [15].

# 5-Experimental results
## 5.1 Metrics significance
In section 4 we have defined a set of metrics that reflects the nature of the application functions. The interpretation of these metrics can be done as follows:

$\gamma$: this metric indicates the average parallelism of a function. The designer can refer to $\gamma$ to classify the functions and to focus on those that have the largest values. Indeed those having the largest values offer more optimization opportunities since they are likely to present a number of implementation alternatives offered by their inherent parallelism. Scheduling a function within a short time constraint will lead to the exploitation of its parallelism while scheduling the same function within a large time constraint will lead to a decrease in the exploitation of its parallelism. By using a multi-time constrained scheduler [16] it is possible to generate time vs. resources trade-off curves on which the points represents implementation alternatives. These curves are very valuable for the designer since he can use them to make some architectural choices (no parallelism-> software implementation, high parallelism->hardware implementation).

MOM: this metric indicates the frequency of global memory accesses, i.e., accesses to input/output data and to memory levels "above" the register level. By referring to this metric the designer can see which functions require special care for implementation: those with large MOM values are most likely to require a good data bandwidth. The MOM metric also indicates the potential need for a memory hierarchy since this metric is computed for all the hierarchy levels of a function graph.

COM: this metric indicates the frequency of "true" control operations, i.e., tests that cannot be eliminated at compile time for example. The designer can refer to this metric to evaluate the need for complex control structures to implement a function. For example functions with high COM values are most likely to better implemented on a GPP processor rather than on a DSP processor since the latter is not well suited for control-dominated algorithms. It also indicates that implementing such functions in hardware would require rather large state machines.
By using the information provided by the metrics, the designer is guided in his architectural choices since he gets an insight of the application's functions. In what follows we illustrate these concepts on some real-life examples.

## 5.2 Analysis result
In this section we give analysis results of some multimedia applications.

### 5.2.1 Smart camera
Our real-life case study is a smart camera. The system to be implemented consists of a video camera associated to a processing unit. The processing unit is responsible for performing an object motion detection on the video streaming from the CMOS sensor of the camera. This smart camera is typically used for monitoring applications such as counting people in subways, tracking car traffic and industrial production lines. The object motion detection consists of several functions such as the detection of the background image, image labelling and other typical video processing algorithms: threshold, dilatation, erosion,... An example of an executing application is shown in Figure 4.
In this study we have characterized the functions of the application as done in the previous example, but we also have performed the system estimation using the tool Design Trotter [2]. The processing part of the smart camera application is composed of 31 functions which represent a total of 1470 lines of C code. The estimation
has been performed rapidly, with computation times comprised between 5 minutes for the most complex functions (some functions include up to 200 sub-graphs) and less than 1 minutes for the simplest functions on a PIII-700Mhz PC. The overall estimation has been performed in less than 2 hours, this demonstrates the value of the method in the case of large design spaces. Firstly, Figure 5 shows the characterization step results. The designer can use these results to classify the functions and to imagine a potential implementation target.
The first observation which can be made is that all the functions have high MOM values, (0.72 on average, more than 2 operations out of 3), this is due to the fact that there are numerous reads of data from the video stream and that the application is highly hierarchical (nested loop structures for example) and that the DFGs are rather short. This implies that this application requires either a big local memory (data reuse) or high-end I/O mechanisms (parallel data reading/writing).
Next we observe that $\gamma$ values are very different, from 1.27 for Convolvetabhisto up to 43.8 for TestGravity. By using these values, it is possible to sort the functions and to find out in what order they should be estimated. Focusing on the most critical ones first enables to sketch an appropriate architecture and also to take reusing into account: the resources allocated to the most critical functions may be reused for the less critical. Finally COM values are comprised between 0 and 0.3 which denotes that control is not dominant, like in the previous example this is justified by the fact that most of the tests in the application are deterministic.

The functions TestGravity and Label have a big γ value, so they have big potential of acceleration obtained by a dedicated Hardware circuits. In the contrary, Add and Sub have small γ value. For such functions there is no need for dedicated hardware accelerator. Those two functions can be realized by the processor since they involve standard mathematical functions (addition and subtraction).

From this preliminary analysis, many architecture conclusions can be made. The first one is that this application can be efficiently mapped on a RISC processor equipped with a high performance bus like a ARM-7 or the LEON processor which support the AMBA bus. Figure 6 shows a first architecture design with two accelerators connected to the AMBA bus.

The hierarchical description (based on the HCDFG graph) of the media function allows analysis at various hierarchical levels. γ value shown in Figure 5 is a global value that indicates the potential speed-up for the function TestGravity seen as a single bloc. It is obtained by a combination of all the subgraphs that constitute this function. A more detailed analysis can be done to estimate those different subgraphs individually. This finer analysis can provide different possible implementations for the function TestGravity. Indeed, it is possible that TestGravity contains subgraphs with a high γ value (>>1) and other with low γ value (near 1) and the combination gives a high γ. In this case, it is not necessary to implement the whole TestGravity as a hardware accelerator. Such granularity analysis examples are given in the following DWT and DCT examples.

### 5.2.2 DWT / DCT transform
If some image compression capabilities are needed, the DWT or DCT functions are required.

### 5.2.2.1 The DCT transform
Figure 7 shows the analysis result for the DCT transform. We give results for the 1D DCT function that is used twice for rows and lines computing and we perform the 2D DCT transform analysis.

We notice that γ has the same value in both 1D and 2D DCT transform. We conclude that –in this example- the parallelism is not affected by the granularity since the data dependency between the raw and the column processing doesn't enable any spatial parallelism. However, one can also observe that firstly the COM value is null since the DCT transform doesn't require any test operation. Secondly, the computation of the inter-iteration (CDFG) available parallelism indicates that a complete loop unrolling can be performed (the unroll factor equals the loop bounds). It means that there is no backward dependence, so pipeline architecture is possible. Finally, an optimized 2D DCT can be realized with two 1D DCT hardware accelerators.

The speed up can be doubled with pipeline if the memory cost is acceptable, in such a case a 8*8 pixels buffer is required between 1D modules. Note that memory requirements can be extracted from the HCDFG with the fast Balasa method [3].

### 5.2.2.2 The DWT transform
The second example is the DWT transform. Figure 8 shows its analysis result. We consider three hierarchical levels, the first one is the analysis of each function in the 1D DWT. The second one is the analysis of the whole 1D DWT (horizontal and vertical). The third level is the analysis of the whole 2D DWT. We notice that γ is increasing between the first and the second level and remains unchanged in the third level. Those results leads to propose two hardware accelerators for the second level (1D DWT(H) and 1D DWT (V)). As COM values are null further improvement can be obtained using pipeline technique.

### 5.2.3 Example of delay/cost IP estimation for the hardware projection of the TestGravity function
Once a critical function has been detected and identified as promising for a hardware implementation our framework enables to rapidly obtain a delay / cost tradeoff curves through the architectural projection step. This function is composed of 378 C code lines, translated into 2408 lines of HCDFG. The corresponding graph is made of 200 sub-graphs. The results obtained for system-level estimation are presented in Figure 9. As firstly indicated by the γ metric this function has a good speedup potential. We have chosen not to show all the possible solutions since the number of resources required for very high speedup factor was extremely high. The most expensive solution shown permits a speedup factor of almost 12 using an architecture enable to perform simultaneously 11 ALU like operations + 7 multiplications and 43 data R/W. On the other hand the cheapest solution only requires one operation of each type at a time but requires a longer execution time.

Figure 10 shows the results of the hardware projection of three solutions onto the Xilinx V400EPQ2 FPGA. The solutions selected are solution 21 (no speedup), solution 11 (speedup=2.05) and solution 1 (speedup=11.85). For each solution the estimated execution time is given (in ns) as well as the estimated number of Logic Cells (LC) and Dedicated Cells (DC).

# 6 Conclusion

In this paper, we have presented an approach for the conception of multimedia processor. This approach is based on two steps. Firstly, multimedia applications are characterised, which results in a set of metrics. These metrics are computed using a hierarchical graph-based representation of the application in order to point out the application hot spots in terms of memory bandwidth, processing parallelism and relative control/processing/data transfers influence at each level of granularity. The second step consists in building a parametrizable multimedia library which is adapted to the application requirement according to the metrics results. Experiences with typical image processing algorithms show firstly how functions with a high potential of optimization can be detected and secondly how the characterization can finely highlight architectural opportunities and directions to improve application-architecture matching. The development of embedded devices gives an extra challenge, since these devices have in general a small energy budget. Without significant energy reduction techniques and energy saving architectures, battery life constraints will limit the capabilities of these devices and reduce seriously their autonomy. Energy reduction can be done at various architectural levels using various techniques ranging from hardware optimizations at low level (transistor or gate level) to power management software approaches (scaling voltage and frequency to reduce power). It also can be driven at algorithmic level. Our approach is currently being extended to support the second case.

## References

[1] Azzedine A. A fast exploration of real-time scheduling & multi-granularity HW/SW solutions for power/area tradeoffs, UBS/LESTER Tech. Report, Jan. 2004.
[2] Auguin M, Ben Chehida K, Diguet JPh, Fornari X, Fouilliart AM, Gamrat C, Gogniat G, Kajfasz Ph, and Le Moullec Y. Partitioning and CoDesign tools & methodology for Reconfigurable Computing: the EPICURE philosophy. Proceedings of the Third International Workshop on Systems, Architectures, Modeling Simulation (SAMOS03), July 2003, Samos, Greece.
[3] Balasa F, Cathoor F. Background memory area estimation for multi-dimensional signal processing systems, IEEE Trans. On VLSI systems 1995;3 (2): 157-172.
[4] Berekovic M, Pirsch P, Selinger T, Miro C, Lafage A, Wels K, Heer C, Ghigo G, "Architecture of an Image Rendering Co-Processor for MPEG-4 Visual Compositing, Kluwer Journal of VLSI Signal Processing Systems 2002; 31(2): 157-171.
[5] Carro L, Kreutz M, Wagner F and Oyamada M. System Synthesis for Multiprocessor Embedded Applications. DATE'00, Paris, France, 2000
[6] Diguet J-Ph., Sentieys O, Philippe J-L and Martin E. Probabilistic Resource Estimation for pipeline architecture. IEEE Work. on VLSI Signal Processing, Sakai, Japan, oct. 1995
[7] Gove R J. The MVP: a highly-integrated video compression chip. Proceedings of IEEE Data Compression Conference, March, 28-31 1994, Snowbird, Utah, pp. 215—224
[8] www.gaisler.com
[9] Guerra L, Potkonjak M and Rabaey J. System-Level Design Guidance Using Algorithm Properties. IEEE Work. on VLSI Signal Processing, San Diego, USA, oct. 1994.
[10] Suzuki K. A 2000-MOPS Embedded RISC Processor with a Rambus DRAM Controller. IEEE journal of solid-state circuits 1999; 34(7).:1010-1021.
[11] Sang H, Myungjim K and Keun-Bae K. Modular and efficient architecture for H.263 video codec VLSI ISCAS 2002 proceedings volume V- pp125, 128.
[12] Ing-Jer H, Wen-Kai H, Rui-Ting G and Chung-Fu K. A cost effective multimedia extension to arm7microprocessors, ISCAS 2002 proceedings, volume II pp 304-307.
[13] Lingfeng L, Danian G and Yun H. A Single chip real time programmable video signal processor
IEEE International Symposium on Circuits and Systems ISCAS 2002 Arizona USA, 26 –29 May 2002.
[14] Maalej I, Gogniat G, Abid M., Philippe J.L. Interface Design Approach For System On Chip Based On Configuration, IEEE International Symposium on Circuits and Systems, ISCAS 2003, Bangkok, Thailand, 25-28 May, 2003.
[15] Le Moullec Y., Ben Amor N., Diguet J-Ph, Philippe J-L and Abid M. Multi-granularity metrics for the era of strongly personalized SOCs. Design Automation and Test in Europe Conference DATE 2003, Munich, Germany 3-7 March, 2003.
[16] Le Moullec Y., Diguet J-Ph., Heller D. and Philippe J-L., "Fast and Adaptive Data-flow and Data-transfer Scheduling for Large Design Space Exploration" ACM/SIGDA GLSVLSI 2002, April 18-19, 2002, New-York, USA.

[17] Nguyen D, Davare A., Orshansky M, Chinnery D, Thompson B, and Keutzer K. Minimization of Dynamic and Static Power Through Joint Assignment of Threshold Voltages and Sizing Optimization. Proceedings of the International Symposium on Low Power Electronics and Design, 2003 (ISLPED'03). pp. 348-353.

[18] Sciuto D, Salice F, Pomante L and Fornaciari W. Metrics for Design Space Exploration of Heterogeneous Multiprocessor Embedded Systems. 10[th] Int. Symp. on H/S Codesign, Estes Park, USA, may 2002.

[19] Vahid F, Gajski D .Closeness metrics for system-level functional partitioning. EDAC'95,Brighton, U.K., sep. 1995

[20] Wuytack S, Diguet J.Ph., Catthoor F. and De Man H. Formalized methodology for data reuse exploration for low-power hierarchical memory mappings. IEEE Transactions on VLSI Systems 1998; 6(4): 529-537.

[21] Bilavarn S, Gogniat G, Philippe J.L. Fast Prototyping of Reconfigurable Architectures: An Estimation And Exploration Methodology from System-Level Specifications. Eleventh ACM International Symposium on Field-Programmable Gate Arrays Monterey, California, February 23-25 2003

Figures Captions:

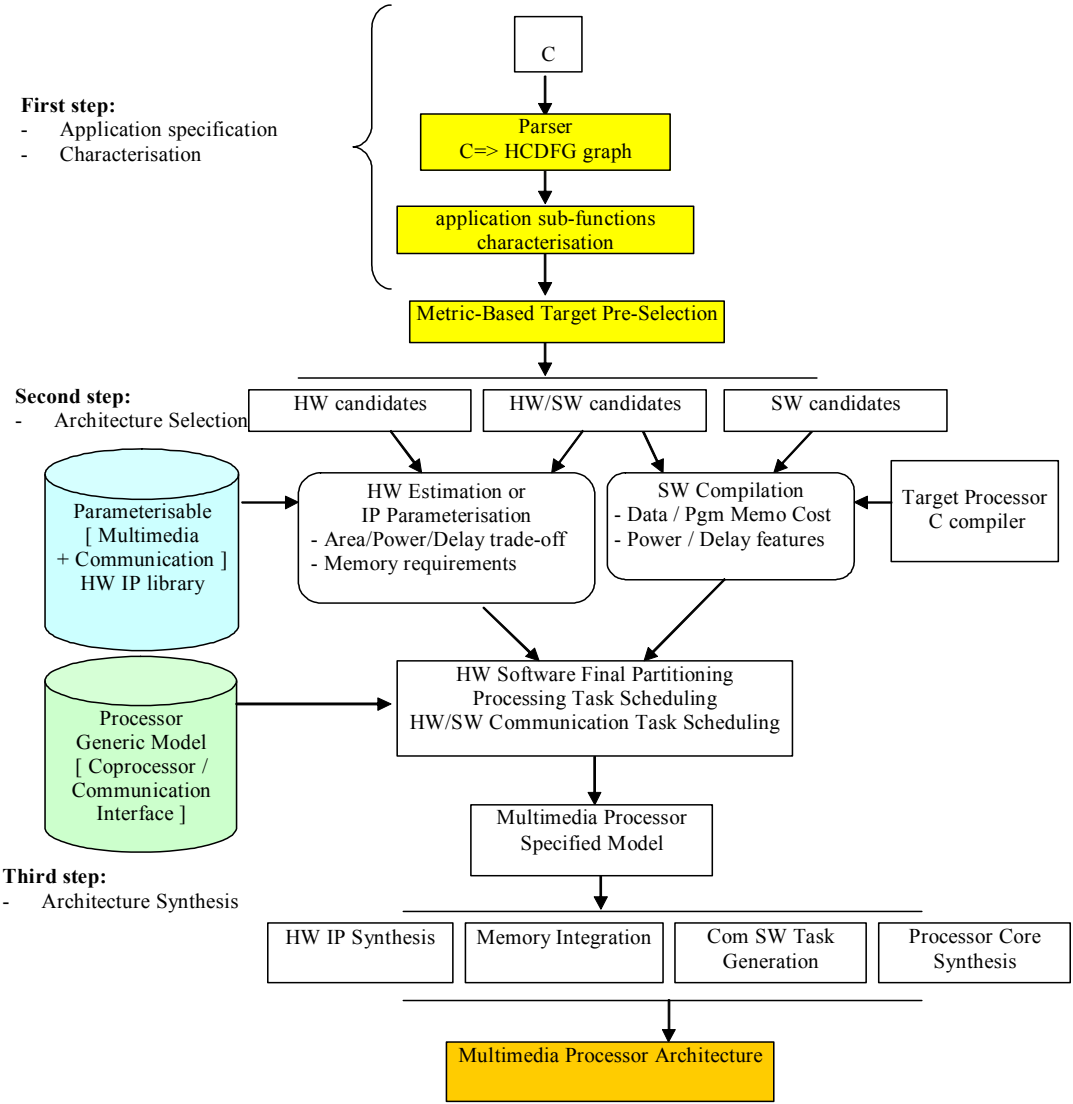

Figure 1 : Metric Based Design Flow

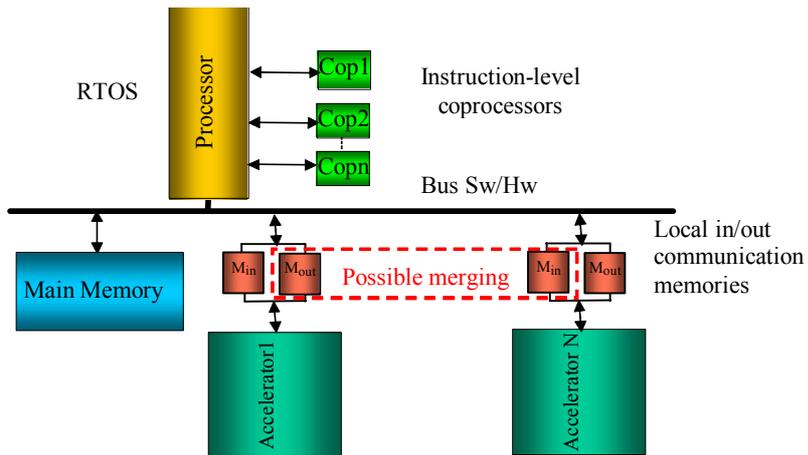

Figure 2: Generic Architecture Model

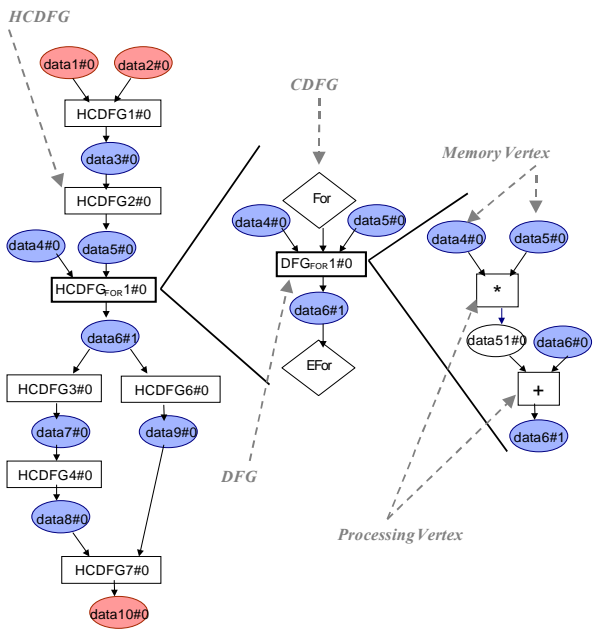

Figure 3 : HCDFG structure

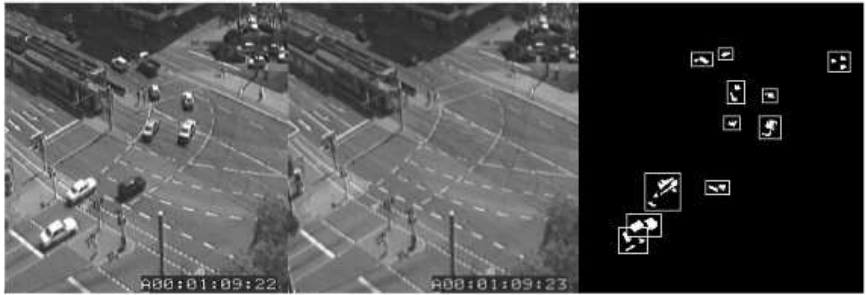

Figure 4 : Smart camera executing. Left: video from the camera, center: background detection, right: moving objects detection

| Function name | γ | MOM [0,1] | COM [0,1] |
|---|---|---|---|
| TestGravity | 43,88 | 0,78 | 0,22 |
| Label | 10,31 | 0,74 | 0,07 |
| ChangeBackground | 5,62 | 0,76 | 0,03 |
| RconstDilat | 4,75 | 0,65 | 0,32 |
| DilatBin | 4,69 | 0,70 | 0,02 |
| HistoThreshold | 4,00 | 0,64 | 0,29 |
| Envelop | 3,91 | 0,66 | 0,13 |
| Absolute | 2,60 | 0,71 | 0,08 |
| ThresholdAdapt | 2,20 | 0,75 | 0,08 |
| ConvolveTabHisto | 1,27 | 0,70 | 0,03 |
| Div | 1,25 | 0,73 | 0,00 |
| GetHistogram | 1,22 | 0,75 | 0,00 |
| SetValue | 1,14 | 0,78 | 0,00 |
| Add | 1,11 | 0,75 | 0,00 |
| Sub | 1,11 | 0,75 | 0,00 |
| ErodBin | 1,10 | 0,73 | 0,01 |

Figure 5 : Smart camera characterization

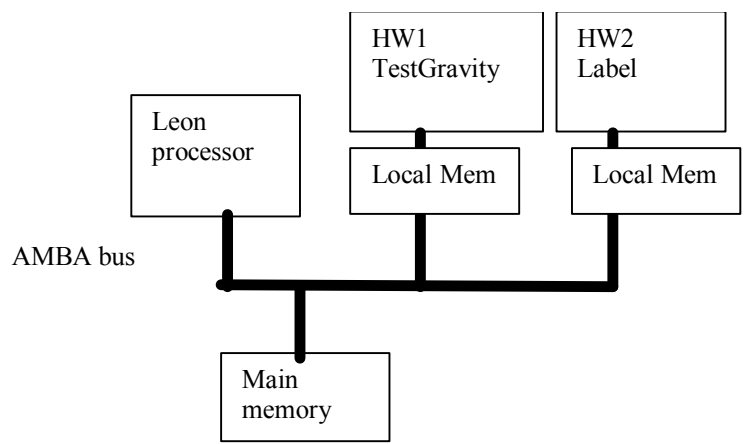

Figure 6 : Proposed Smart Camera Architecture

| Functional bloc | MOM | COM | γ | Max unroll factor |
|---|---|---|---|---|
| DCT8L | 0,58 | 0 | 5,7 | 8 |
| DCT8C | 0,58 | 0 | 5,7 | 8 |
| DCT8x8 | 0,58 | 0 | 5,7 | 8 |

Figure 7 : DCT2D 8x8 metrics

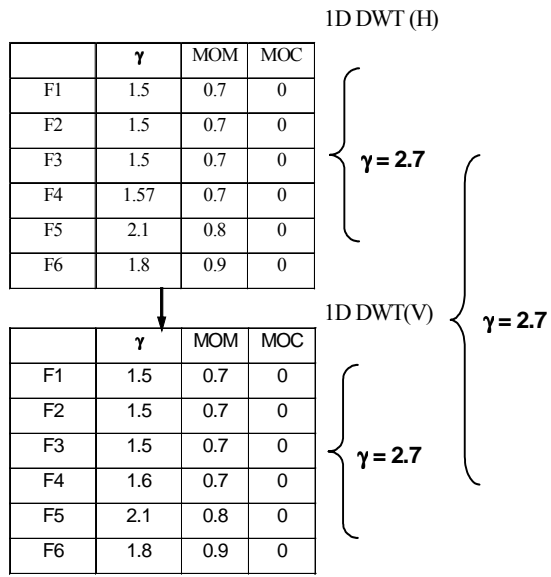

Figure 8: DWT exemple : Lifting scheme

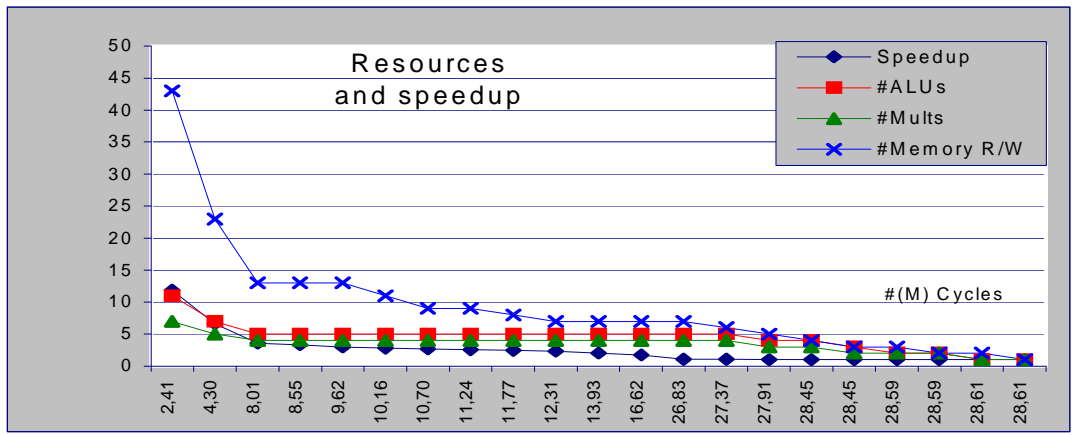

| Solution number | #Cycles | Speedup | #ALUs | #Mults | #Memory R/W |
|---|---|---|---|---|---|
| 1 | 2414976 | 11,85 | 11 | 7 | 43 |
| 2 | 4302848 | 6,65 | 7 | 5 | 23 |
| 3 | 8009992 | 3,57 | 5 | 4 | 13 |
| 4 | 8547816 | 3,35 | 5 | 4 | 13 |
| 5 | 9623464 | 2,97 | 5 | 4 | 13 |
| 6 | 10161288 | 2,82 | 5 | 4 | 11 |
| 7 | 10699112 | 2,67 | 5 | 4 | 9 |
| 8 | 11236936 | 2,55 | 5 | 4 | 9 |
| 9 | 11774760 | 2,43 | 5 | 4 | 8 |
| 10 | 12312584 | 2,32 | 5 | 4 | 7 |
| 11 | 13926056 | 2,05 | 5 | 4 | 7 |
| 12 | 16615176 | 1,72 | 5 | 4 | 7 |
| 13 | 26833832 | 1,07 | 5 | 4 | 7 |
| 14 | 27371656 | 1,05 | 5 | 4 | 6 |
| 15 | 27909480 | 1,02 | 4 | 3 | 5 |
| 16 | 28447500 | 1,01 | 4 | 3 | 4 |
| 17 | 28447503 | 1,01 | 3 | 2 | 3 |
| 18 | 28592958 | 1,00 | 2 | 2 | 3 |
| 19 | 28592960 | 1,00 | 2 | 2 | 2 |

Figure 9 TestGravity trade-off curve

| Solution number | Time (ns) | Nb LC | Nb DC |
|---|---|---|---|
| 1 | 46178247 | 296 | 36 |
| 11 | 266932793 | 253 | 29 |
| 21 | 547212227 | 200 | 18 |

Figure 10 TestGravity projection on Xilinx V400EPQ2

| Metric name | Type of Graph | Formula |
|---|---|---|
| γ | General definition | (1) $\gamma = \dfrac{Nb\ of\ global\ memory\ accesses\ and\ processing\ operations}{Critical\ Path}$ |
| γ | IF graph | (2) $\gamma = P_{true} * \dfrac{Nop_{true}}{CP_{true}} + P_{false} * \dfrac{Nop_{false}}{CP_{false}} + \dfrac{Nop_c}{CP_c}$ |
| γ | Combination of sequential graph | (3) $\gamma_{equivalent} = \dfrac{\sum_{i}^{total\ subgraphs} Nop_i}{\sum_{i}^{total\ subgraphs} CP_i}$ |
| γ | Combination of parallel Graph | (4) $\gamma_{equivalent} = \dfrac{\sum_{i}^{total\ subgraphs} Nop_i}{Max_i\{(CP_i)\}}$ |
| MOM | General definition | (5) $Mom = \dfrac{Nb\ of\ global\ memory\ accesses}{Nb\ of\ processing\ operations + Nb\ of\ global\ memory\ accesses}$ |
| COM | General definition | (6) $Com = \dfrac{Nb\ of\ test\ operations}{Nb\ of\ processing\ operations + Nb\ of\ global\ memory\ accesses}$ |

Table 1 metric definition